\documentstyle[eqsecnum,aps,epsf]{revtex}

\def\beqa{\begin{eqnarray}}
\def\eeqa{\end{eqnarray}}
\def\be{\begin{equation}}
\def\ee{\end{equation}}
\begin{document}
\title{Period-doubling bifurcation in strongly anisotropic Bianchi I quantum 
cosmology}
\author{Michael Bachmann\thanks{mbach@physik.fu-berlin.de, 
http://www.physik.fu-berlin.de/\~{}mbach} 
and Hans-J\"urgen Schmidt\thanks{hjschmi@rz.uni-potsdam.de,
http://www.physik.fu-berlin.de/\~{}hjschmi}}
\address{Institut f\"ur Mathematik, Universit\"at Potsdam,
PF 601553, D-14415 Potsdam, Germany,\\
Institut f\"ur Theoretische Physik, Freie Universit\"at Berlin,
Arnimallee 14, D-14195 Berlin, Germany}
\date{\today}
\maketitle
\begin{abstract}
We solve the Wheeler-DeWitt equation for the 
minisuperspace  of a cosmological model of Bianchi type I 
with a minimally coupled massive scalar field $\phi$ as source 
by generalizing the calculation of Lukash and Schmidt~[1]. 
Contrarily to other approaches we allow strong anisotropy. 
Combining analytical and numerical methods, we apply an adiabatic
approximation 
for $\phi$, and  as  new feature we  find a  period-doubling bifurcation.
This bifurcation takes place near the cosmological quantum boundary, i.e.,
the 
boundary of  the quasiclassical region with oscillating $\psi$-function where 
the WKB-approximation is  good. The numerical calculations suggest that  
 such a  notion of a ``cosmological quantum boundary''   is well-defined,
because sharply beyond that boundary, the WKB-approximation is no more 
 applicable at all.
This result confirms the  adequateness of 
the introduction of  a cosmological quantum boundary in quantum cosmology. 
\end{abstract}

\section{Introduction}
\setcounter{equation}{0} 
The idea to consider the whole Universe as one quantum system has 
already attracted many researchers. One of the approaches to give this 
idea a physical meaning is to consider the superspace $\cal S$ of
all possible spatial 3-geometries together with the matter field degrees of 
freedom as supercoordinates, and the supermomentum space $\cal S^\star$
is formed from the  corresponding second fundamental forms together
with the momenta of the matter fields. 
The dynamics will then be defined by the Einstein field equation
accompanied by the matter field equations. In the last step, this
system will be quantized.

Of course, one has to deal with an infinite number of degrees of
freedom already from the geometric part. The, up to now only, 
way to tackle such a system is to linearize in almost all degrees
of freedom and to take the non-linearities only from the remaining 
finitely many ones. But even this system is almost 
untractable. So, the idea of the minisuperspace arose: One restricts to
a finite-dimensional space of spatial geometries and to matter fields
with finitely many components, disregarding or even ignoring 
all other degrees of freedom. 

From the first glance one could believe that such a big 
simplification would lead to a picture which has nothing to do 
with the real Universe's evolution. However, the results given 
in the last years are encouraging: The dynamics of the minisuperspace
can be written as an equivalent mechanical system, and then the
Schr\"odinger equation for this system carries the name
Wheeler-DeWitt equation which can be solved for simple systems and
give already a surprisingly good picture of the evolution, even for the
case that the geometric part is restricted to only one degree
of freedom, the ``radius''  of the Universe. This corresponds to the
isotropic Friedmann Universe model. For this case,  
closed-form solutions for the Wheeler-DeWitt equation exist. 

In the ordering with respect to simplicity, the next possible geometry is the
Bianchi type I model, where the spatial inner geometry is flat, but the 
expansion is allowed to be anisotropic. For this case, we have 3 degrees of 
freedom for the geometry: the expansion rates into the 3 spatial 
directions.\footnote{Strictly speaking, the axially symmetric Bianchi type 
I model, where 2 of the 3 scale factors coincide, is even simpler, but as one
will see from the calculations: restricting to axial symmetry does not
really simplify the procedure.} 
For this case, the Wheeler-DeWitt equation 
is already quite complicated, so one uses an approximation to solve it.
One of the most powerful of these approximation schemes is the 
Wentzel-Kramers-Brillouin (WKB)-approximation. In Ref.~[1], the first 
order WKB-approximation for the massive scalar field in the 
Bianchi type I model has been deduced, but -- as far  as we are aware --  the 
higher order WKB-approximations have not yet been calculated up to now for
this model.  The analogous calculations for Bianchi type IX have been done by 
Amsterdamski [2]. In both cases, the anisotropy degrees of freedom had been 
assumed to be small.  In the present  paper  we allow also large
anisotropies.  Ref. [3] deals with a similar model allowing large
anisotropies, too; 
 however, the authors of [3] use a  simplicial minisuperspace model, so the
results 
are not directly comparable to ours. 

\bigskip

The text is organized as 
 follows: Sect. II presents the geometry of the Bianchi type I cosmological
model,
 Sect. III the corresponding Wheeler-DeWitt equation and  Sect. IV its solutions
 both analytically and numerically. 
 Sect. V shortly reviews Refs. [6 - 42], i.e., the 
earlier developments of the topic, and  discusses the results.

 In the appendix, we show how the different versions to  solve  the 
factor-ordering problem influence  the solutions of the Wheeler-DeWitt
equation.

\section{The Bianchi type I model}
The metric of a spatially flat cosmological model is deduced as
follows: One requires that an abelian 3-dimensional 
isometry group acts transitively on 3-dimensional spatial
hypersurfaces which consequently have to be flat 3-spaces. 
This restricts the possible metrics to
\be
\label{dist1}
ds^2= N^2 dt^2 - g_{i j}dx^{i} dx^{j}
\ee
where $N=N(t)$ is the lapse function which can be put to $N=1$ 
by a time reparametrization, $i, j = 1,2,3$, and 
$g_{i j}$ represents a symmetric positive definit matrix whose 
components depend on $t$ only. So we have 6 free components from the
first consideration. However, restricting to metrics being a solution of
Einstein's field equation we can simplify as follows: At any initial time,
say $t=0$, we can choose the initial condition
\be
\label{metric1}
g_{i j} = \delta_{i j}= \mbox{diag}(1,1,1) 
\quad {\rm at} \quad  t=0
\ee
without loss of generality, and then by a spatial rotation not
changing Eq.~(\ref{metric1}), the second fundamental form, i.e.  
$d g_{i j}/dt$, can be brought into diagonal form. 
By use of the Einstein equation one can show that under these
circumstances, $g_{i j}$ will keep its diagonal form for
all times. Thus: The Bianchi type I model containing non-diagonal 
terms does {\it not} represent a generalization, and we can say: 
Without loss of generality let
\be
\label{metric2}
g_{i j} = \mbox{diag}(A^2 (t), B^2 (t),C^2 (t))
\ee
with certain positive functions $A,B,C$.

After an obvious rearrangement of the terms we get now from 
Eqs.~(\ref{dist1}) and (\ref{metric2})
\be
\label{inv2}
ds^2= N^2 dt^2 - e^{2\alpha} \left(
e^{2s+2\sqrt 3 r}dx^2 +
e^{2s-2\sqrt 3 r}dy^2 +
e^{-4s}dz^2 \right)
\ee
where $\alpha, r,s$ represent 3 arbitrary real functions depending on $t$.
The Hubble parameter is the mean expansion, i.e. with a dot denoting 
$d/dt$ we get 
\[ H=\frac{1}{3} \left( \frac{\dot A}{A} + \frac{\dot B}{B} + 
\frac{\dot C}{C}  \right)\].
In the notation of Eq.~(\ref{inv2}), $H=d\alpha /dt$ is the Hubble parameter, 
and we restrict our considerations to the expanding Universe, i.e., 
to $H>0$. 
The model goes over to the spatially flat isotropic Friedmann model if the
dimensionless anisotropy parameter
\be
\label{aniso1}
\eta = \frac{1}{H}\sqrt{\dot r^2 + \dot s^2}
\ee
tends to zero.

It is essential to observe that this approach leading to the
Wheeler-DeWitt equation breaks the Lorentz invariance of the
system already at this level of geometry: The 3+1 decomposition of 
space-time is already made from the beginning. The remaining 
degrees of freedom within the geometry are the following: 
$\tilde t(t)$ as arbitrary function as long as $N(t)$ remains
unspecified, and $\tilde t(t) = \pm t$ after having fixed $N=1$. 
Permutations of the 3 spatial coordinates, and for the axially
symmetric case also spatial rotations: If $r=0$, then a rotation in the
$xy$-plane is an additional symmetry. We presented this 
geometric part so explicitly, because contradicting
statements exist about this behavior in the literature; 
example: In our interpretation,
$ds^2=dt^2-dx^2-dy^2-dz^2$ and $ds^2=dt^2- t^2 dx^2-dy^2-dz^2$
represent different geometries inspite of the fact that they are locally
isometric space-times. 

We use a one-component real massive scalar field $\phi$ as source, 
and we interpret it as follows: Either it can really be such a scalar field
(i.e. a spin zero field), 
e.g. a Higgs field which dominated the early Universe's evolution
but disappeared after a symmetry breaking effect. Or, it mimics
any realistic matter field in a region where spin is negligible.
Let $m$ denote that mass of the scalar field. For the GUT-mediated 
inflationary model, one considers the following order of magnitude:
$m \approx 10^{-5} m_{\rm Pl}$, where $m_{\rm Pl}$ is the Planck 
mass.\footnote{In  units with $c=1$, the Planck mass is about $10^{19}$ GeV.}

Up to now, we did not restrict the spatial coordinates $x$, $y$ and $z$.
At the classical level (i.e. for locally solving the Einstein field equation),
 it makes no difference, whether they cover all the reals or whether they 
are cyclic ones. But already for global classical considerations it {\it
makes}
 a difference: Supposed,    $x$, $y$ and $z$ are cyclic, then even for 
the case $r=s\equiv 0$ the metric is not spatially isotropic, because the 
natural length of a closed spatial geodesic depends on the  chosen direction. 
 However, we get the essential difference only after quantization. 
 If nothing different is said, we assume  
$x$, $y$ and $z$ to be cyclic with modulus 1, i.e. $x$ and $x+1$ 
 represent the  same point etc.

\section{The Wheeler-DeWitt equation}
The Wheeler-DeWitt equation is always a zero energy
Schr\"odinger equation due to time-reparametrization invariance 
of the gravitational action. However, this is not really a problem,
because the Schr\"odinger equation with non-vanishing energy can be
brought to zero-energy form by a suitable shift of the potential, 
see e.g. Ref.~[4] for a class of transformations of this type. 
     
In the present paper, we do not consider the entropy of the system;  
we only want to mention  the recently published result of Kleinert [5] 
on how the incorporation of  entropy can change the model.

To formulate the equations that lead to the Wheeler-DeWitt equation [6] 
as easily as possible, it proves useful to 
apply units such that $\hbar = c= 4\pi G/3=1$. Then the Lagrangian
of the   minimally coupled scalar field in Einstein's theory of gravity  
 (see the first of Refs.~[1] for details) is given by  
\be
\label{lag}
L=L_g+L_m,
\ee
where  the gravitational part is
\be
\label{lag:g}
L_g = \frac{1}{2N}e^{3\alpha} \left( 
 \dot r^2+\dot s^2 - \dot\alpha ^2
\right) 
\ee
and the matter Lagrangian is
\be
\label{lag:m}
L_m = \frac{1}{2N}e^{3\alpha} \left( 
 \dot \phi^2 - m^2 \phi ^2 N^2 
\right) .
\ee
Here we supposed the lapse function $N$ to be constant in time.
The minisuperspace is 4-dimensional, its coordinates are 
$q^\mu=(q^0, q^1,q^2,q^3)$ 
with $q^0  =\alpha $, $q^1 = \phi $, $q^2 =r $, and $q^3 = s$. 
The usefulness of the Misner parametrization 
Eq.~(\ref{inv2}) of
the anisotropy is obvious: The kinetic part of the Lagrangian is already in 
diagonal form. By defining the minisuperspace metric 
$f_{\mu\nu}=e^{3\alpha}\,\mbox{diag}(-1,1,1,1)$ the Lagrangian (\ref{lag})
is compactly rewritten as
\begin{equation}
  \label{lag:ms}
  L=\frac{1}{2N}\left(f_{\mu\nu}\dot{q}^\mu\dot{q}^\nu -
m^2 (q^1)^2  N^2 e^{3q^0}\right).
\end{equation}
Thus the introduction of canonical momenta 
\begin{equation}
\label{canmom}
p_\mu=\frac{\partial L}{\partial \dot{q}^\mu}
=\frac{1}{N}f_{\mu\nu}\dot{q}^\nu
\end{equation}
allows us to write down the classical Hamiltonian of the minisuperspace
\begin{equation}
  \label{ham}
  {\cal H}=\frac{N}{2}\left(p_\mu p^\mu+
m^2      (q^1)^2   e^{3q^0}      \right)=0.
\end{equation}
Time-reparametrization invariance implies $\frac{\partial   {\cal H}}{
\partial N} =0$, i.e.,   ${\cal H}=0$. 
The quantization in coordinate space is done as usual by 
going over to the appropriate operators $p_\mu = - i \hbar \partial
/\partial q^\mu$
with the partial derivative $\partial$.

\bigskip

From now on, we will choose two different  versions for  
the operator ordering in  $p_\mu p^\mu$: 

\medskip

First variant, variant A, which is the  most simple one, 
follows the interpretation in Ref.~[1]  
that the solutions of the 
forthcoming Wheeler-DeWitt equation will not essentially depend on 
this choice. 

\medskip

Second, variant B, solves the factor ordering problem by
 applying  the supercovariance principle: It says, that 
 the essential equations have to be covariant ones with respect to the
 (mini-)superspace metric. Taking variant B, we have to replace
$p_\mu$ by  $p_\mu = - i \hbar \nabla /\nabla q^\mu$
 where $\nabla$ denotes the supercovariant derivative\footnote{Here 
we apply the fact that the metric is always covariantly constant, so
different 
orderings give the same result.} 
defined 
by $f_{\mu \nu}$.

\medskip

To cover both variants simultaneously, we introduce a parameter
$\epsilon$ as follows: $\epsilon=0$ for variant A and $\epsilon=1$
for variant B. Accordingly, we denote the two variants of the
 D'Alembert operator\footnote{The minus sign in eqs. (3.7/3.8) is inserted
to compensate for the factor $i^2$ in front of $p_\mu p^\mu$.}
  $\Box_\epsilon$ by
\begin{equation}
  \label{box}
  \Box_0 = 
-f^{\mu\nu}\frac{\partial^2}{\partial q^\mu \partial q^\nu}.
\end{equation}
and 
\begin{equation}
  \label{box2}
  \Box_1 = 
-f^{\mu\nu}\frac{\nabla^2}{\nabla q^\mu \nabla q^\nu}.
\end{equation}
For the world function $\psi(q^\mu)$ we get the relations
\be
\Box_1 \psi =  \Box_0  \psi + 3 \psi_{,0} e^{-3\alpha}
\ee
where\footnote{If the dimension of the minisuperspace is $D$ instead of
$4$, then the
factor $3$  in front of $\psi_{,0}$in Eq. (3.9) will be replaced by
$3(D-2)/2$, so 
$\Box_0$ and $\Box_1$ coincide for $D=2$.} 
\be
 \Box_0 \psi=e^{-3\alpha}\left[\psi_{,00}- \psi_{,11} - \psi_{,22}
- \psi_{,33} \right]
\ee
The stationary Schr\"odinger equation with zero energy in minisuperspace
 $  \hat{\cal H}\psi(q^\mu)=0$  
with the wave function $\psi(q^\mu)$ and the 
Hamilton operator $\hat{\cal H}$ is called
{\em Wheeler-DeWitt equation}
and possesses the form
\begin{equation}
  \label{wdw1}
  \hat{\cal H}\psi(q^\mu)=\frac{N}{2}\left(\hbar^2  \Box_\epsilon + 
m^2 (q^1)^2 e^{3q^0} \right)\psi(q^\mu)
=0
\end{equation}
for the Bianchi type I Universe (\ref{inv2}). 
 In details, 
we get after dividing by $Ne^{3\alpha}/2$
\be
0=\left(  \hbar^2 \left[\frac{\partial^2}{\partial \alpha^2} + 
3 \epsilon  \frac{\partial}{\partial \alpha} -
\frac{\partial^2}{\partial \phi^2} -
\frac{\partial^2}{\partial r^2}- \frac{\partial^2}{\partial s^2}
\right] + m^2 \phi^2 e^{6 \alpha}    \right) \psi(\alpha, \phi, r, s).
\ee
In what follows we derive an approximation scheme to solve the
Wheeler-DeWitt equation.

\section{Solutions of the Wheeler-DeWitt equation}
Before we go over  to solve Eq. (3.12) we want to clarify in subsection A 
what happens if we neglect the anisotropy from the beginning. In subsections
B to D we analytically obtain solutions in a well-defined approximation, and
 in subsection E we present the results of our numerical calculation. 
\subsection{Isotropic models}
For variant A, i.e., $\epsilon =0$ in Eq. (3.11) we  get
\be
0=\left(\hbar^2 \left[\frac{\partial^2}{\partial \alpha^2} -  
\frac{\partial^2}{\partial \phi^2}
\right] + m^2 \phi^2 e^{6 \alpha}    \right) \psi (\alpha,\phi)
\ee
i.e., we simply remove all $r$- and
$s$-dependencies. For variant B, however, this recipe does not work because
the 
covariant derivatives mix  the coordinates. After a short calculation one
finds out
that here also Eq. (4.1) is the correct Wheeler-DeWitt equation. The reason
for this coincidence 
is as follows: The 2-dimensional superspace is conformally flat, and the 
D'Alembertian is conformally invariant, cf. the second paper in Ref. 1 for
more
details.
\bigskip

Defining $a = e^\alpha>0$ as new variable we have to replace
$\frac{\partial}{\partial \alpha}$ by $a \frac{\partial}{\partial a}$, and
we get
 from Eq. (4.1) now 
\be
0=\left(\hbar^2 \left[  a \frac{\partial}{\partial a}a
\frac{\partial}{\partial a} - 
\frac{\partial^2}{\partial \phi^2}
\right] + m^2 \phi^2 a^{6}    \right) \tilde \psi (a,\phi)
\ee
 a version of the Wheeler-DeWitt equation favoured in Ref. [7]. 
\subsection{Separation ansatz}
Now we look for the  solutions of Eq. (3.12). First, we make a separation
ansatz
\be
\psi(\alpha,\phi,r,s) = \chi (\alpha,\phi) \cdot \rho(r) \cdot \sigma(s).
\ee
It turns out that $\rho_{,22}/\rho =c_2$ and $\sigma_{,33}/\sigma =c_3$
represent constants, i.e., we get
$$
\rho (r) = \rho_1 \exp(\sqrt{ c_2}\  r) +  \rho_2 \exp(- \sqrt {c_2} \  r)
$$
and
$$
\sigma (s) = \sigma_1 \exp(\sqrt{ c_3} \  s) +  \sigma_2 \exp(- \sqrt {c_3}
\  s)
$$
with arbitrary constants $\rho_i$ and $\sigma_i$. For negative 
values $c_2$, the function $\rho(r)$ has a sinus-function behaviour.

One can interpret these cases as follows: If both $c_2$ and $c_3$ 
are negative, then we have plane waves in $r$- and $s$-direction.
 This is in agreement with the fact, that a translation into
 $r$- or  $s$-direction can be compensated by a coordinate transformation 
 in metric (2.4)  
(i.e., by multiplying $x$, $y$, and $z$ with suitable constants), so that
all $r$- and $s$-values should be equally probable. However, this 
interpretation works only in case that we allow  $x$, $y$, and $z$ to
 cover all the reals. If we, however, restrict  $x$, $y$, and $z$ to be 
 cyclic coordinates, then this argument in favour of equal 
distribution of the $r$- and $s$- values is no more valid. 

If, on the contrary, one of the constants   $c_2$ or  $c_3$ is  
non-negative, then (up to singular exceptions) 
 the product $\rho(r) \cdot \sigma(s)$ tends to
$\pm \infty$  as $r$ or $s$ does. This allows the following interpretation: 
The probability to have small anisotropy is exponentially small. 

For the moment we keep the further interpretation open and
continue the calculation. We define the
constant $c_1$ via $c_1 +c_2 +c_3=0$ and insert Eq. (4.3) into (3.12); we
get:  
\be
0=\left(\hbar^2 \left[\frac{\partial^2}{\partial \alpha^2} 
+ 3 \epsilon \frac{\partial}{\partial \alpha}  -  
\frac{\partial^2}{\partial \phi^2}
\right] + c_1 \hbar^2 + m^2 \phi^2 e^{6 \alpha}    \right) \chi (\alpha,\phi).
\ee
This equation is for variant A just the Wheeler-DeWitt equation (4.1),
however, 
now not the zero-energy equation but the equation with energy proportional to
 the constant $c_1$ representing the anisotropic degrees of freedom. 

We
 redefine the coordinate $\alpha$ to $v=a^3=e^{3\alpha}$. The 
coordinate $v$ is proportional to the spatial volume. Then Eq. (4.4) 
goes over to (after dividing by 9) 
\be
0=\left(\hbar^2 \left[v \frac{\partial}{\partial v} v
\frac{\partial}{\partial v} 
+  \epsilon v  \frac{\partial}{\partial v}  -  
\frac{\partial^2}{\partial \phi^2}
\right] + c_1 \hbar^2/9  + m^2 \phi^2 v^2/9    \right) \tilde \chi (v,\phi).
\ee
As one can see in comparison with Eq. (4.2): Instead of the $a^6$ term we
now have a $v^2$-potential.  
\subsection{Adiabatic scalar field}
In this subsection, we assume the scalar field $\phi$ to be almost constant.
 Then $\gamma$ defined by $\gamma^2 = m^2 \phi^2 /(9  \hbar^2)$ is an
 adiabatic constant. 
In Eq. (4.5) we omit the $\phi$-derivative corresponding to the adiabatic
approach.
To this end we consider the following equation for the exact 
dependence on $v$ of  a  new function $ \hat \chi$
\be
\left(\hbar^2 \left[v \frac{\partial}{\partial v} v
\frac{\partial}{\partial v} 
+  \epsilon v  \frac{\partial}{\partial v}  
\right] + c_1 \hbar^2/9  + \gamma^2  v^2 \hbar^2    \right) \hat \chi (v,\phi)
 = E(\phi) \cdot \hat  \chi (v,\phi),
\ee
 where $E(\phi)$ is an yet undetermined eigenvalue; here, $\phi$ plays the
role of 
a parameter only. We define a new adiabatic constant $\Lambda$
 by $\Lambda^2 = E(\phi) \hbar^{-2} - c_1 /9 $. Then Eq. (4.6) becomes for
 $\epsilon =0$
\be
\left( v \frac{\partial}{\partial v} v \frac{\partial}{\partial v} 
  + \gamma^2  v^2 - \Lambda^2    \right) \hat \chi =0.
\ee
Eq. (4.7) is a Bessel type ordinary differential equation. If we replace in
Eq.~(4.7) $\gamma v$ by $x$ and $ \hat \chi (v,\phi)$ by $y(x,\phi)$
 we get  exactly Bessel's form  
\be
x^2 \  \frac{d^2 y}{dx^2}
   \ + \ x \ \frac{dy}{dx} \  + \  (x^2-\Lambda^2) \  y  \  =  \  0,
\ee
whose solutions are (the constants $C_i$ depend on $\Lambda$)

\be
y=C_1  J_\Lambda(x)  +C_2J_{-\Lambda}(x) \  .
\ee
The $J_\Lambda$ are called cylinder or Bessel functions and have the
development
\be
  J_\Lambda(x) = \sum^\infty_{k=0}(-1)^k \left(\frac{x}{2}
\right)^{\Lambda+2k}
  \left( k!   \right)^{-1}   \left( \Gamma(\Lambda +k+1)  \right)^{-1}.
\ee
Now we make the ansatz\footnote{If only a discrete set of
values $\Lambda$ appear,
Eq. (4.11) will be replaced 
 by an analogous sum.}
\be
\tilde \chi(v,\phi)=\int_\Lambda b_\Lambda(\phi) \hat\chi_\Lambda(v,\phi)
d\Lambda,
\ee
where we assigned to $\hat\chi$ the subscript $\Lambda$ to emphasize
$\hat\chi_\Lambda$
being an eigenfunction with eigenvalue $\Lambda$. Thus the general solution
of Eq.~(4.7)
is written as
\be
 \hat\chi_\Lambda(v,\phi) = C_1 J_\Lambda(\gamma v)
 +  C_2 J_{-\Lambda}(\gamma v).
\ee
Inserting the ansatz (4.11) into (4.5) we get
\be
0 = \int_\Lambda - \hbar^2 
\frac{\partial^2}{\partial \phi^2} \left[   b_\Lambda(\phi)
\hat\chi(v,\phi)   \right]
+    b_\Lambda(\phi)  \left[
 \hbar^2 \left(v \frac{\partial}{\partial v} v \frac{\partial}{\partial v} 
+ \frac{c_1}{9}  \right) + \frac{m^2}{9} v^2 \phi^2
 \right]   \hat\chi(v,\phi) d \Lambda.
\ee
In the second term of Eq. (4.13) we replace the term in the brackets $[$ $]$ 
by the eigenvalue
$E(\phi)$ according to Eq. (4.6). $E(\phi)$ depends on $c_2$, $c_3$ and
$\Lambda$ only, i.e., there is no $\phi$-dependence, and we write $E$ instead
of $E(\phi)$ 
\be
0 = \int_\Lambda - \hbar^2 
\frac{\partial^2}{\partial \phi^2} \left[   b_\Lambda(\phi)
\hat\chi(v,\phi)   \right]
+    b_\Lambda(\phi) \ E \ 
   \hat\chi(v,\phi) d \Lambda.
\ee
In Eq.~(4.14) we first write out the $\frac{\partial^2}{\partial \phi^2}$
 applied to the product, but we assume that only 
$\frac{\partial^2}{\partial \phi^2}   b_\Lambda(\phi)$ is essential, and
disregarding the other terms\footnote{This is usually called
adiabatic approximation with respect to $\phi$.}
 and assuming that the $\hat \chi$-s are all independent 
we obtain the simple differential equation
\be
\left[ - \hbar^2  \frac{\partial^2}{\partial \phi^2} + \  E \ \right]
b_\Lambda(\phi)=0
\ee
which possesses solutions
\be
   b_\Lambda(\phi) = B_1 e^{\sqrt E \, \phi/\hbar} 
+ B_2 e^{- \sqrt E \, \phi/\hbar}.
\ee
Finally, we get the general wave function
\begin{eqnarray}
\label{wave1}
\psi^{(\phi)}_{c_2\,c_3}(\alpha,\phi,r,s)&=&
\left(
 \rho_1 \exp\left\{\sqrt{ c_2}\  r\right\} +  \rho_2 \exp\left\{- \sqrt
{c_2} \  r\right\} \right)
 \left(
 \sigma_1 \exp\left\{\sqrt{ c_3} \  s\right\} +  \sigma_2 \exp\left\{-
\sqrt {c_3} \  s\right\}
\right)\nonumber\\
&&\times\int_\Lambda
\left[
B_1 \exp\left\{\sqrt{\Lambda-(c_2+c_3)/9} \, \phi\right\} 
+ B_2 \exp\left\{- \sqrt{\Lambda-(c_2+c_3)/9} \, \phi\right\}
\right] \nonumber\\
&&\times\left[
C_1 J_\Lambda\left(m\phi e^{3\alpha}/3\hbar\right)
 +  C_2 J_{-\Lambda}\left(m\phi e^{3\alpha}/3\hbar\right)
\right]
d\Lambda 
\end{eqnarray}
with real and for the moment continuous eigenvalues $c_2, c_3$. The
superscript $( \phi ) $
 at $\psi$ 
indicates that we have treated the scalar field adiabatically. This
solution has to be 
specified by appropriate choices for the constants $\rho_1$, $\rho_2$,
$\sigma_1$, $\sigma_2$,
$B_1$, $B_2$, $C_1$, and $C_2$ from boundary conditions. 
Before discussing the result~(\ref{wave1}), however, we
calculate the wave function by considering the scale factor $\alpha$ as the
adiabatic 
variable.
\subsection{Adiabatic Scale Factor Approach}
Another adiabatic method to solve the Wheeler-DeWitt equation~(4.4) is to
treat the scale factor
$\alpha$ as a slowly varying variable. 

Once more, we reexpress the scale factor $\alpha$ by $v\equiv e^{3\alpha}$,
implying that 
we consider the Wheeler-DeWitt equation~(4.5) in the following.    
Considering $v$ as the adiabatic variable means that we neglect derivatives
with respect 
to $v$ in a first step. Then Eq.~(4.5) becomes
\begin{equation}
\label{sfa00}
\left( - \hbar^2\frac{\partial^2}{\partial \phi^2}
 + \frac{\hbar^2}{9}\, c_1 + \frac{1}{9}\,m^2 \phi^2 v^2    \right) \hat
\chi (v,\phi) 
= E(v)\, \hat \chi (v,\phi)
\end{equation}
with $E(v)$ being a still undetermined eigenvalue. By defining 
\begin{equation}
\label{sfa01}
\omega = \frac{1}{3}\,m v, \quad \tilde x =
\sqrt{\frac{\omega}{\hbar}}\,\phi, \quad
\eta=\frac{1}{\hbar\omega}\,\left(E(v)-\frac{\hbar^2}{9}\,c_1\right),
\end{equation}
the differential equation~(\ref{sfa00}) is simply
transformed into that for the dimensionless harmonic oscillator
\begin{equation}
\label{sfa02}
\left[\frac{d^2}{d \tilde{x}^2}+(\eta-\tilde{x}^2) \right] \tilde y(\tilde
x) = 0.
\end{equation}
By substituting $x=\tilde{x}^2$ and then transforming $\tilde
y(x)=e^{-x/2}\,y(x)$, 
we eventually obtain Kummer's differential equation
\begin{equation}
\label{sfa03}
x\frac{d^2 y(x)}{d x^2}+(\mu-x)\,\frac{d y(x)}{d x}-\nu\,y(x)=0,
\end{equation}
where in our case $\mu=1/2$ and $\nu=(1-\eta)/4$.
This differential equation possesses the general solution
\begin{equation}
\label{sfa04}
y(x) = A\; _1F_1(\nu,\mu;x)+B\; _1F_1(\nu-\mu+1,2-\mu;x)\,x^{1-\mu},
\end{equation}
expressed with the help of the confluent hypergeometric function
$_1F_1(a,b;x)$. $A$ and $B$
are constants. Thus the partial problem of finding the solution of
Eq.~(\ref{sfa00}) is done
and the result is written as
\begin{equation}
\label{sfa05}
\hat\chi_\eta(v,\phi)=e^{-\omega(v)\phi^2/2\hbar}
\left[A\; _1F_1\left((1-\eta)/4,1/2;\omega(v)\phi^2/\hbar\right)+
B\, \sqrt{\frac{\omega(v)}{\hbar}}\phi\;
_1F_1\left((3-\eta)/4,3/2;\omega(v)\phi^2/\hbar\right)\right]
\end{equation}
with $\hat{\chi}_\eta$ denoting the eigenfunction to the real, continuous
eigenvalue $\eta$. 

For further processing to find an adiabatic solution for the Wheeler-DeWitt
equation~(4.5)
we make the integral ansatz
\begin{equation}
\label{sfa06}
\tilde{\chi}(v,\phi)=\int_\eta g_\eta(v)\,\hat\chi_\eta(v,\phi)\,d\eta.
\end{equation}
Inserting this into Eq.~(4.5), we obtain (for $\epsilon=0$)
\begin{equation}
\label{sfa07}
0=\int_\eta \hbar^2 v \frac{\partial}{\partial v}v
\frac{\partial}{\partial v}
\left[g_\eta(v)\,\hat\chi_\eta(v,\phi) \right]+g_\eta(v)
\left[-\hbar^2\frac{\partial^2}{\partial \phi^2} 
+\frac{1}{9}\,(\hbar^2c_1+m^2\phi^2v^2)\right]\hat\chi_\eta(v,\phi)\,d\eta.
\end{equation}
The expression enclosed in the brackets $[$ $]$
 in the second term is substituted by the eigenvalue
\begin{equation}
\label{sfa08}
E(v)=\frac{1}{9}\left[3\hbar m \eta v-\hbar^2(c_2+c_3)\right],
\end{equation}
following from Eq.~(\ref{sfa00}) and the definitions (\ref{sfa01}). 
In contrast to the preceding approach with
an adiabatic scalar field, $E(v)$ 
depends explicitly on the adiabatic variable which is in this case $v$.
Note that $E(v)$ also 
depends on the eigenvalues $c_2$, $c_3$, and $\eta$.
Now we utilize the assumption of adiabaticity with respect to $v$ by 
neglecting terms containing derivatives $\partial\hat\chi_\eta/\partial v$
and 
$\partial^2\hat\chi_\eta/\partial v^2$ which appear in the first term of
Eq.~(\ref{sfa07}). 
Furthermore supposing that all functions 
$\hat\chi_\eta(v,\phi)$ are independent with respect to $\eta$, each
integrand in 
Eq.~(\ref{sfa07}) vanishes. Thus we remain with solving the differential
equation
\begin{equation}
\label{sfa09}
\left[\hbar^2 v \frac{\partial}{\partial v}v  \frac{\partial}{\partial
v}+E(v) \right]g(v)=0
\end{equation}
or, more explicitly,
\begin{equation}
\label{sfa10}
\left[v^2 \frac{\partial^2}{\partial v^2}+v\frac{\partial}{\partial
v}+\kappa_1v-
\frac{\kappa_2^2}{4}\right]g(v)=0 
\end{equation}
with $\kappa_1=m\eta/3\hbar$ and $\kappa_2^2=4(c_2+c_3)/9$. Applying the
transformation
$x=2\sqrt{\kappa_1 v}$ and denoting the solution as $y(x)=g(v(x))$,
Eq.~(\ref{sfa10}) takes 
the same Bessel form as given in Eq.~(4.8), whereas the index is $\kappa_1$
instead of
$\Lambda$
here. For this reason the general solution of Eq.~(\ref{sfa10}) reads
\begin{equation}
\label{sfa11}
g_\eta(v)=D_1\, J_{4(c_2+c_3)/9}\left(\sqrt{4m\eta v/3\hbar}\right)+
D_2\, J_{-4(c_2+c_3)/9}\left(\sqrt{4m\eta v/3\hbar}\right)
\end{equation}
with constants $D_1$ and $D_2$.

Thus, the complete general wavefunction obtained with an adiabatic scale
factor approach
(indicated by superscript~$\alpha$) is found to be
\begin{eqnarray}
\label{sfa12}
\psi^{(\alpha)}_{c_2\,c_3}(\alpha,\phi,r,s)&=&\left(\rho_1
\exp\left\{\sqrt{ c_2}\  r\right\} +  
\rho_2 \exp\left\{- \sqrt {c_2} \  r\right\} 
\right)\left(\sigma_1 \exp\left\{\sqrt{ c_3} \  s\right\}) +  
\sigma_2 \exp\left\{- \sqrt {c_3} \ s\right\}\right)\nonumber\\
&&\times\int_\eta\left[D_1\, J_{4(c_2+c_3)/9}\left(\sqrt{4m\eta
e^{3\alpha}/3\hbar}\right)+
D_2\, J_{-4(c_2+c_3)/9}\left(\sqrt{4m\eta e^{3\alpha}/3\hbar}\right)\right]
\exp\left\{-m\phi^2e^{3\alpha}/6\hbar\right\} \nonumber\\
&&\times 
\left[A\; _1F_1\left((1-\eta)/4,1/2;m\phi^2e^{3\alpha}/3\hbar\right)+
B\, \sqrt{\frac{m}{3\hbar}}\phi e^{3\alpha/2}\;
_1F_1\left((3-\eta)/4,3/2;m\phi^2e^{3\alpha}/3\hbar\right)\right]d\eta
\end{eqnarray}
with constants $\rho_1$, $\rho_2$, $\sigma_1$, $\sigma_2$, $A$, $B$, $D_1$,
and $D_2$
to be determined via appropriate boundary conditions. The anisotropy
quantum numbers
$c_2$ and $c_3$ are supposed to be real at this stage. Certain boundary
conditions,
such as requiring the wavefunction to vanish for infinitely large values of
$\alpha$ and/or
 $- \alpha$, 
lead to replacing the integral by a sum over $\eta$. 
\subsection{Visualization of the world function}
\begin{figure}[t]
\centerline{
\setlength{\unitlength}{1cm}
\begin{picture}(12.0,9.7)
\put(0,0){\makebox(12,9.7){\epsfxsize=12cm \epsfbox{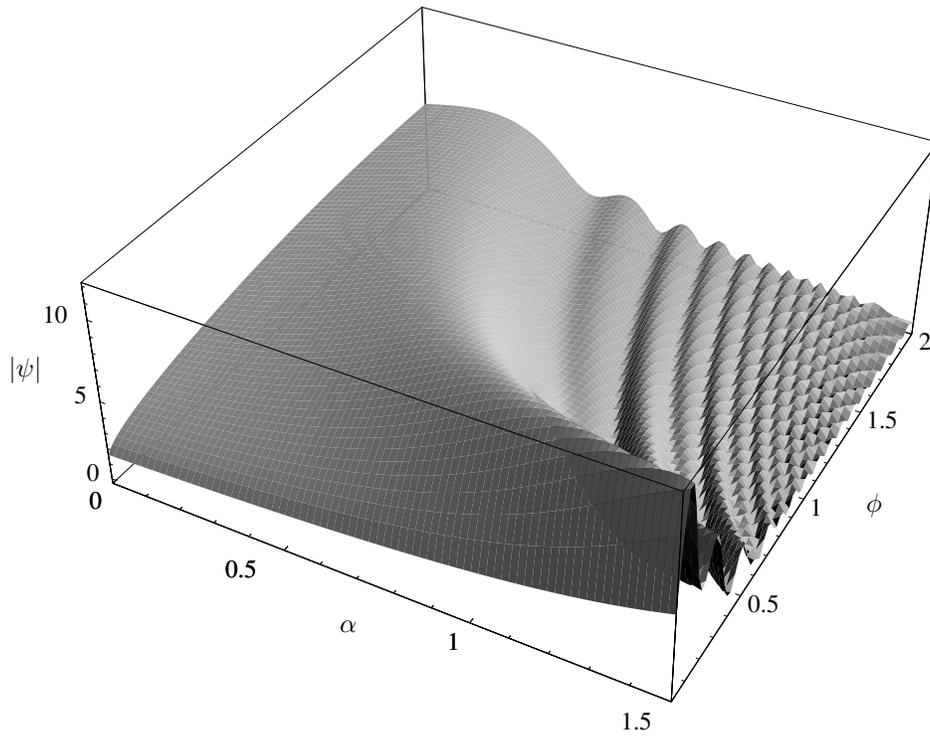}}}
\put(4,1.4){$\alpha$}
\put(11.0,3.0){$\phi$}
\put(-0.4,4.8){$\vert\psi\vert$}
\end{picture}}
\caption{\label{fig1}
The absolute value of the wave function $\vert \psi \vert$
in dependence on $\alpha$ and $\phi$ for parameters $r=s=1$ and $c_2=4$ and $c_3=7$.}
\end{figure}

\begin{figure}[t]
\centerline{
\setlength{\unitlength}{1cm}
\begin{picture}(12.0,9.7)
\put(0,0){\makebox(12,9.7){\epsfxsize=12cm \epsfbox{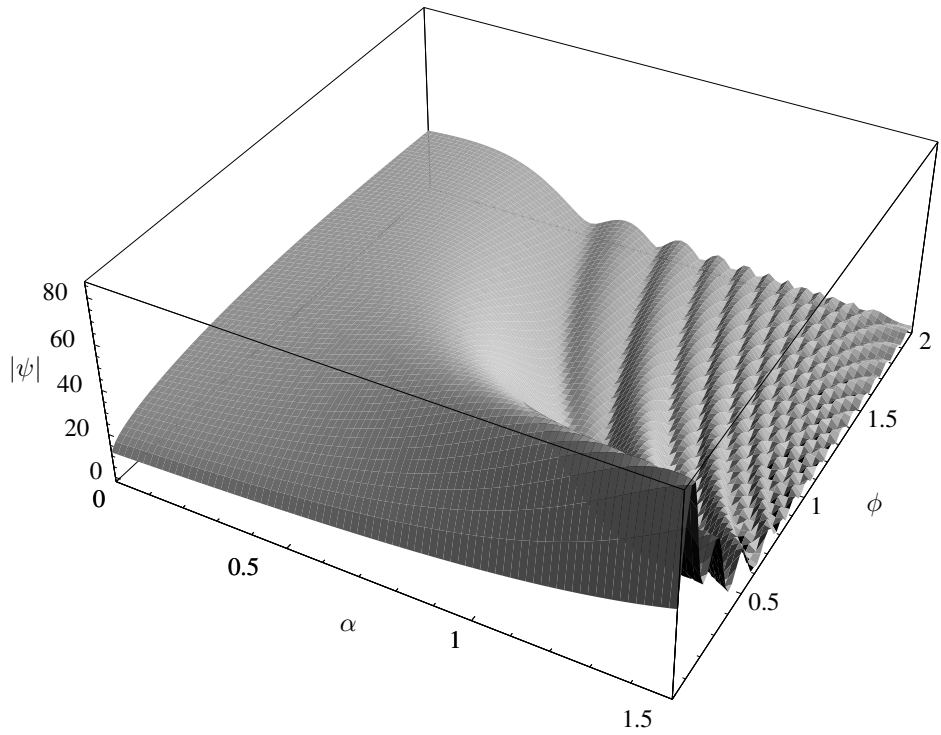}}}
\put(4,1.4){$\alpha$}
\put(11.0,3.0){$\phi$}
\put(-0.4,4.8){$\vert\psi\vert$}
\end{picture}}
\caption{\label{fig2}
The same as Fig.~\ref{fig1}, now $c_2=- 4$ and $c_3=7$. The different
spacing of the  $\vert \psi \vert$-values should be noted. 
}
\end{figure}

Now we visualize the results of our numerical calculation of the world
function (\ref{wave1}). 
To get an impression about the shape of the probability amplitude, we concentrate
on the absolute value of the wave function $\vert \psi \vert$ which is given in these six
diagrams in dependence of $\alpha$, the logarithm of the  cosmic scale  factor, and
the scalar field $\phi$. Each figure is printed at fixed values of the other
parameters in Eq.~(\ref{wave1}). We chose $\rho_1=\sigma_1=B_1=C_2=0$ and 
$\rho_2=\sigma_2=B_2=C_1=1$ to ensure
convergence for small and large values of $\alpha$ and $\phi$. The absolute wave function 
$\vert \psi(\alpha,\phi) \vert$ is not normalized, however, when comparing different figures
(for instance Fig.~\ref{fig1} and Fig.~\ref{fig2}), 
higher function values mean an increased probability density. 
\begin{figure}[t]
\centerline{
\setlength{\unitlength}{1cm}
\begin{picture}(12.0,9.7)
\put(0,0){\makebox(12,9.7){\epsfxsize=12cm \epsfbox{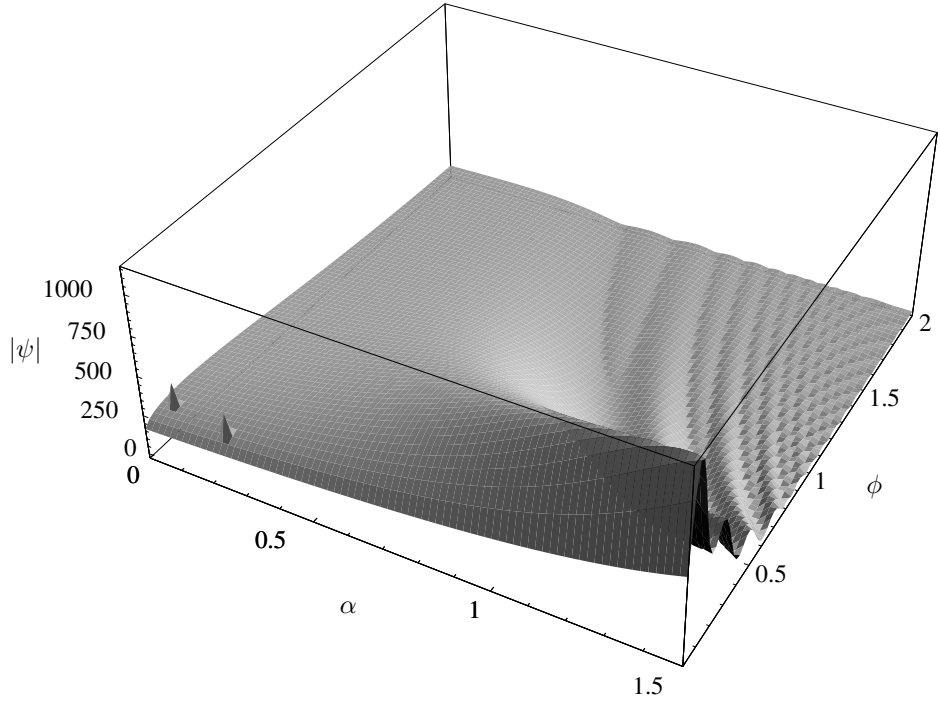}}}
\put(4,1.4){$\alpha$}
\put(11.0,3.0){$\phi$}
\put(-0.4,4.8){$\vert\psi\vert$}
\end{picture}}
\caption{\label{fig3}
The same as Fig.~\ref{fig1}, now $c_2=-4$ and $c_3=-7$. The small
discontinuities in the figure in the lower left corner may be numerical artefacts, so we
do not comment them further. In comparison to Figs.~\ref{fig1} and \ref{fig2}, the different spacing 
of the  $\vert \psi \vert$-axis is still greater.
}
\end{figure}
\begin{figure}[t]
\centerline{
\setlength{\unitlength}{1cm}
\begin{picture}(12.0,9.7)
\put(0,0){\makebox(12,9.7){\epsfxsize=12cm \epsfbox{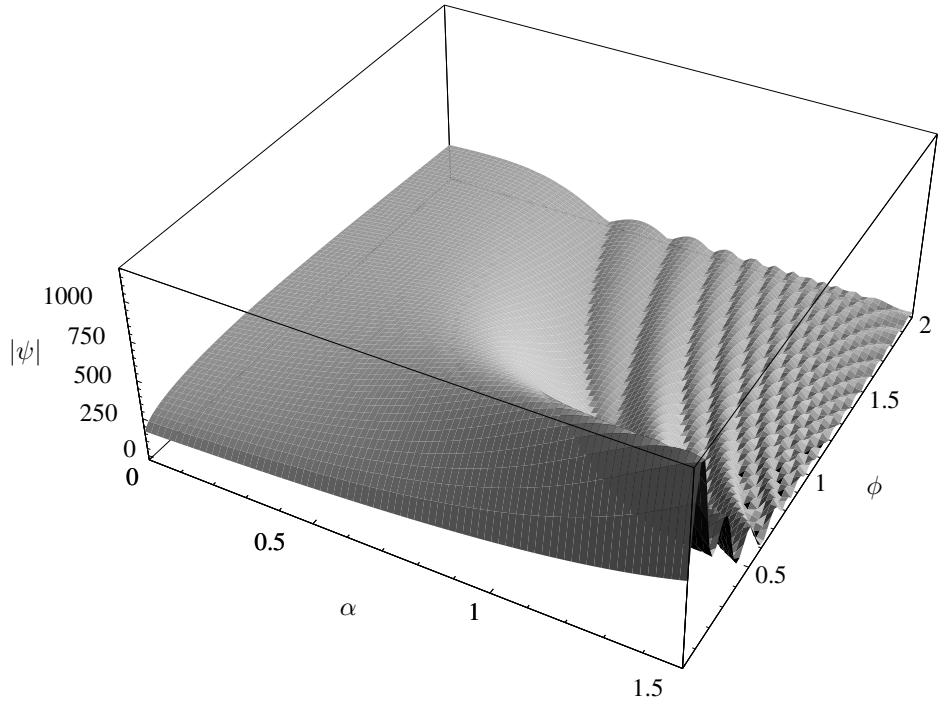}}}
\put(4,1.4){$\alpha$}
\put(11.0,3.0){$\phi$}
\put(-0.4,4.8){$\vert\psi\vert$}
\end{picture}}
\caption{\label{fig4}
The wave function in the isotropic limit, represented by $c_2=0$ and $c_3=0$. 
}
\end{figure}
\begin{figure}[t]
\centerline{
\setlength{\unitlength}{1cm}
\begin{picture}(12.0,9.7)
\put(0,0){\makebox(12,9.7){\epsfxsize=12cm \epsfbox{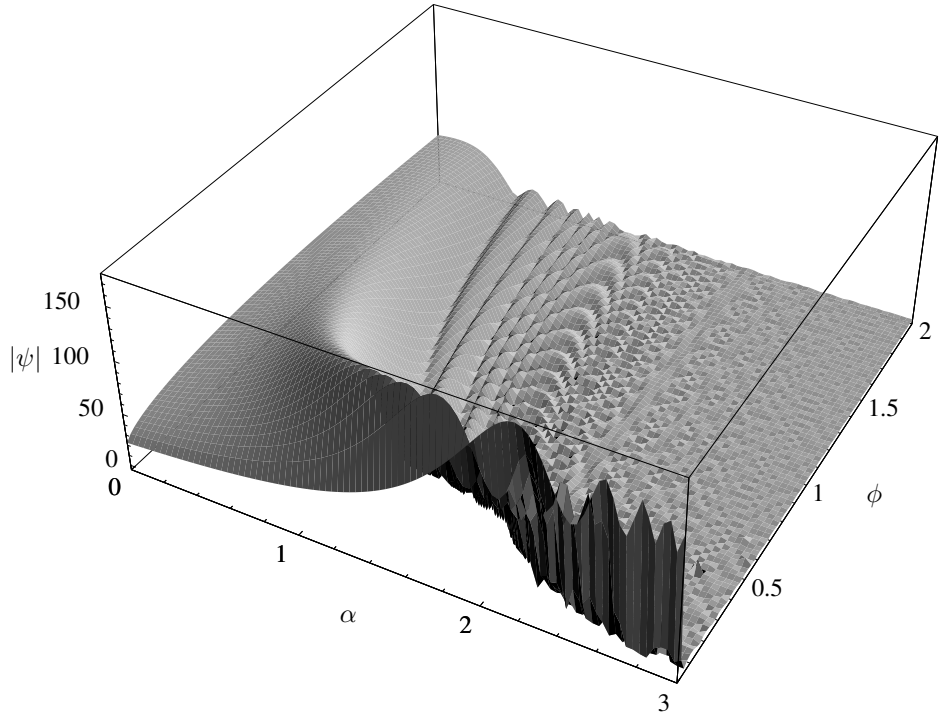}}}
\put(4,1.4){$\alpha$}
\put(11.0,3.0){$\phi$}
\put(-0.4,4.8){$\vert\psi\vert$}
\end{picture}}
\caption{\label{fig5}
The wave function with quantum numbers $c_2=1$ and $c_3=1$. 
This figure is also an example for the influence of the anisotropy 
upon the shape of the wave function.
}
\end{figure}
\begin{figure}[t]
\centerline{
\setlength{\unitlength}{1cm}
\begin{picture}(12.0,9.7)
\put(0,0){\makebox(12,9.7){\epsfxsize=12cm \epsfbox{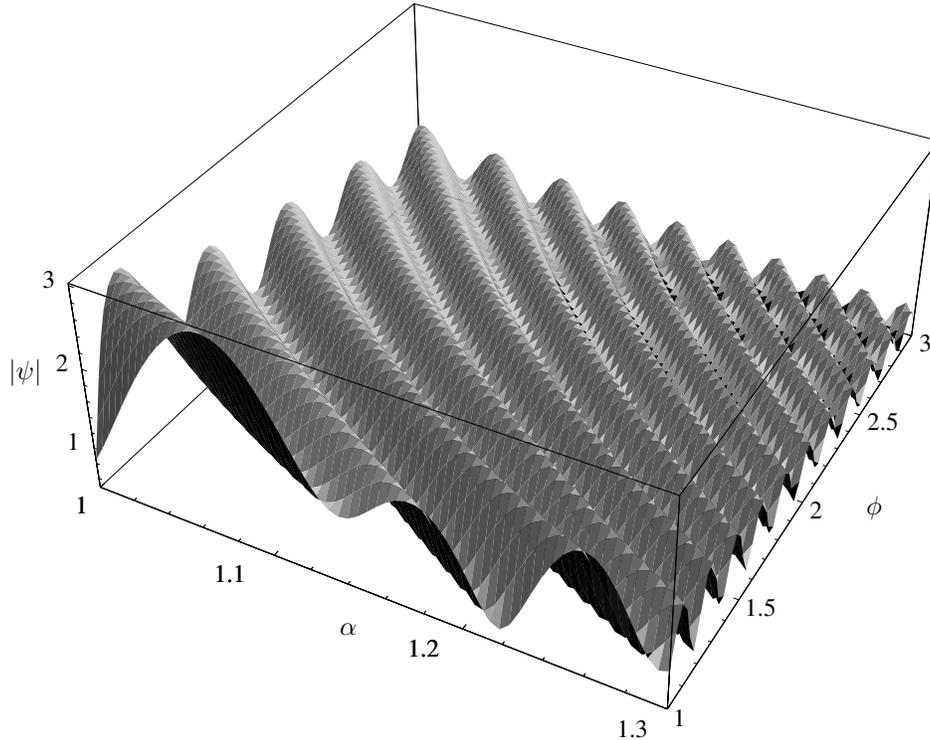}}}
\put(4,1.4){$\alpha$}
\put(11.0,3.0){$\phi$}
\put(-0.4,4.8){$\vert\psi\vert$}
\end{picture}}
\caption{\label{fig6}
Magnification of a subrange of Fig.~\ref{fig1}. From the description it
becomes clear which part of Fig.~\ref{fig1} it represents. We show it here to make
clear, how quickly the WKB-approximation becomes a good approximation. 
}
\end{figure}
For all these different ranges of parameters leading to the results 
shown in Figs.~\ref{fig1}--\ref{fig5} we find essentially the 
same behaviour: The right part of the figure is the region where the
WKB-approximation is valid, that means, the solution is close to a sinus-shaped wave with a
slowly varying amplitude (see Fig.~\ref{fig6}). 
The left hand side is the quantum region where no
oscillations exist at all. 
The range where these two behaviours go into each other, is relatively
sharply defined. This range is what one usually calls the ``cosmological 
quantum boundary'', and these pictures show up that this notion is
relatively well-defined. 

An interesting new feature of these figures is the following: They show a 
period-doubling bifurcation of the frequency if one looks from the bottom 
to the top of these
figures with increasing values of $\phi$. 
\section{Discussion}
Before we discuss the results of the present paper we give a short
review to other papers: 

The original paper [6] by DeWitt was seminal 
 to the whole development of quantum gravity. Its application to
 minisuperspace cosmological models has been developed by several
researchers, 
e.g. the authors and editors of refs. [7] and [8]. At that time, the 
main discussion dealt with the closed isotropic  Friedmann 
models, and the spatially flat Friedmann models had been included as less
interesting limit cases, too.

\medskip

In [9] and [10],
 Sch\"on and Hajicek  perform  a quantization of systems with quadratic
constraints and  discuss the Wheeler-DeWitt equation.  In   ref. 
[11],  Landsman gives details which Hilbert space might 
be appropriate for the isotropic minisuperspace quantization, a topic,
which was 
not much discussed before.

\medskip

In the recent preprint by Kim [12], the minimally coupled massless
scalar field in an open   isotropic Friedmann 
model has been discussed and its Wheeler-DeWitt equation be solved
from the  point of view that the Universe is created quantum mechanically 
from ``nothing''.

\medskip

Recently, Capozziello and Lambiase~[13] investigated the connection between
the Hartle criterion for selecting correlated regions in the configuration
space of 
dynamical variables and an associated Noether symmetry. This relationship
serves
to classify solutions of the Wheeler-DeWitt equations in semi-classical 
minisuperspace models. Thus, the oscillatory behaviour of a subset of
solutions entails 
the presence of Noether symmetries which, in the consequence, select
classical 
Universe models.

\medskip

The paper
 [14] by  Schunck and  Mielke is related to the 
 models discussed here as follows: They apply the 
Wagoner-Bekenstein-Starobinsky-trans\-for\-ma\-tion for
classifying inflationary solutions with scalar field as source, and this
 classification should also apply to the corresponding Wheeler-DeWitt 
equations. 

\medskip

In [15]  and   [16], a  one-parameter set of minisuperspace metrics in
arbitrary dimensions is considered, from which we have chosen 
only that one which gives classically the correct correspondence to
Einstein's theory.
 In [15], the signature of the superspace metric in dependence on the 
 signature of the underlying manifold is evaluated with the result
 that the normal-hyperbolic character of the Wheeler-DeWitt equation exists 
 only for the Euclidean and the Lorentzian signature of the underlying
manifold.   

\medskip

In [17], Horiguchi,  Maeda and  Sakmaoto perform  an expansion
of solutions of the Wheeler-DeWitt equation in powers of the Planck length.
 Vilenkin [18] 
compares several approaches to quantum cosmology, 
 Kim and  Page [19]
discuss quantum Friedmann models and power-law inflation,
 Kim [20] compares  
quantum Friedmann models with conformally and minimally coupled
scalar fields. In Ref. [21], Bleyer and  Ivashchuk 
discuss  multidimensional cosmological models and
 their corresponding Wheeler-DeWitt equations.
 Page [22] solves  
 the Wheeler-DeWitt equation for scalar fields as source.

\medskip

Ref. [23] represents  the famous paper in which the   ``Hartle-Hawking
boundary
conditions" for the Wheeler-DeWitt equation have been derived.
One takes a path integral over all such space-times $V_4$ whose boundary is
the prescribed spatial hypersurface  $V_3$. 

\medskip

In Ref.~[24], Kiefer constructs wavepackets in minisuperspace for a 
Friedmann Universe. An adiabatic approach is used in the
case of a massive scalar field, thereby assuming the 
scale parameter $a$ to be slightly changing only. 
In Zeh [24], these solutions are discussed under the point of view 
 of the definition of the  direction of time. 
 Our Eq. (4.1) coincides with the form discussed in Ref. [24] (Eq. (6.5) of
Zeh, and
Eq. (2.2) of Kiefer). The approximation of our 
 subsection IV D was used by Kiefer~[24] for solving
the Wheeler-DeWitt equation of a Friedmann Universe. 
 In contrast to Kiefer's procedure, where the harmonic oscillator has discrete
eigenvalues and thus the wave function decreases for large values of the scale
factor $\alpha$, we allow all real eigenvalues.

\medskip

Conradi~[25] 
solves the Wheeler-DeWitt equation for Bianchi type IX and a
massive scalar field. 

\medskip

Grishchuk and Sidorov~[26]
discuss the initial conditions for the Wheeler-DeWitt equation,
especially for the massive scalar field in a closed Friedmann
model.

\medskip

In Ref.~[27], Amendola, Khalatnikov, Litterio, and Occhionero
consider quantum cosmology with a complex field.

\medskip

Guendelmann and Kaganovich~[28]
discuss cosmic time in quantum cosmology. The factor-ordering
problem is solved such that the kinetic term gets the form
$\Box + \xi R$ where $\xi \le \xi_{conf}$.

\medskip

A comparison of the minisuperspace of minimally and conformally coupled
scalar fields was done by Page~[29]. He solves the factor-ordering problem
of the
Wheeler-DeWitt equation by requiring that the kinetic term is
proportional to the Laplacian in the minisuperspace metric, i.e., our
variant~B
with $\epsilon=1$. The
classical equation is similar to the geodesic equation in
superspace. 

\medskip

Refs. [30] deal with quantum cosmology from the path integral point
of view: The Wheeler-DeWitt equation can be derived in a first 
approximation from the corresponding  path integral. 
 Halliwell [30] solves the factor-ordering problem by requiring invariance
with
respect to field redefinition of both 3-metric and lapse function.  
 Jafariztadeh,  Darabi and Rastegar [30] apply the method of
 Duru and Kleinert to evaluate the  path integral for  quantum cosmology, 
cf. Kleinert [30]. 

DeWitt [30] assumes the path integral to be the more fundamental approach,
but 
 the Wheeler-DeWitt equation in the minisuperspace to remain a good
 approximation to it.

\medskip

By considering the Wheeler-DeWitt equation
$$\left(a^{-p}\frac{\partial}{\partial a} a^p \frac{\partial}
{\partial a} + a^4 \frac{\Lambda}{3}\right)\psi(a)=0   $$
Gibbons and Grishchuk~[31] obtain the result that inflation is typical in
the set of
spatially flat Friedmann models in Einstein's theory with a
$\Lambda$-term. The parameter $p$ is due to the factor-ordering
ambiguity, they take $p=1$ as a preferred value.
An analogous result is given by Hawking and Page~[32].
Melnikov and Pevcov~[33]
discuss the factor-ordering problem for the Wheeler-DeWitt
equation in closed and open Friedmann models and give some
solutions to it.

\medskip

Reuter and Schmidt~[34] and Schmidt~[35] derive the Wheeler-DeWitt
equation of fourth-order gravity for a spatially flat
Friedmann model and compare with the corresponding conformally
 equivalent (due to the Bicknell theorem) second order models.
The solution of Reuter/Schmidt for flat Friedmann
models is generalized in Ref.~[36] by Pimentel and Obregon
to closed and open models. Fabris and Reuter [36] continue to generalize
 the results of [34] to show that the Bicknell theorem applies also at the
 level of the Wheeler-DeWitt equation.

\medskip

Rainer~[37] gives an overview on higher dimensions and discusses 3 types of 
conformal transformations of different levels for the Wheeler-DeWitt equation.
Quite recently, the solutions of the Wheeler-DeWitt equation in comparison
with the appearance of singularities is treated by Mongan~[38]. 
Zhang and Shen~[39] consider quantum cosmology with a complex 
scalar field at finite temperature. 

\medskip

A critical discussion of the Wheeler-DeWitt equation and an alternative
quantization scheme 
is presented in Ref.~[40] by Peres. 

\medskip

From recent constraint calculations of Hwang and Noh~[41] to an inflation
model 
based on a non-minimally coupled massive 
scalar field and comparisons with observational data ({\it COBE}-DMR), 
one can state that minimal coupling is a good approximation for
inflationary models.
Following Futamase and Maeda~[42], the coupling constant is either quite
small, $\xi<1/1000$, 
or negative. According to the coupling factor $(1-\xi\phi^2)$ in front of
$R$, critical
behaviour appears for $\xi = \phi^{-2}$, i.e. for positive values of $\xi$
only.

\bigskip

Let us now summarize our results:   
We solved the Wheeler-DeWitt equation for the 
minisuperspace  of a cosmological model of Bianchi type I 
with a minimally coupled massive scalar field $\phi$ as source 
by generalizing the calculation of Lukash and Schmidt~[1]. 
Contrarily to other approaches we allowed strong  anisotropy. 

Combining analytical and numerical methods, we  applied  an adiabatic
approximation 
for $\phi$, and  as  new feature we  found a  period-doubling bifurcation
of the 
typical solutions. This bifurcation takes place near the cosmological
quantum boundary, i.e., the 
boundary of  the quasiclassical region with oscillating $\psi$-function where 
the WKB-approximation is  good. The numerical calculations suggest that  
 such a  notion of a ``cosmological quantum boundary''   is well-defined,
because sharply beyond that boundary, the WKB-approximation is no more 
 applicable at all.

This result confirms the  adequateness of 
the introduction of  a cosmological quantum boundary in quantum cosmology. 
We applied  the supercovariance principle, i.e., the underlying theory should
 also  be covariant with respect to transformations representing a  
 mixture between space-time and  matter degrees of freedom. 
 With our figures, we tried to 
vizualize the birth of the Universe.

\section*{Acknowledgement}
One of us (M.B.) is supported by the Studienstiftung des deutschen Volkes.
H.-J. S. gratefully acknowledges financial support from DFG
 (KL 256/31-2).
We thank the colleagues of the Free University Berlin, 
where this work has been done, especially Prof.
H. Kleinert and Dr. A. Pelster, for kind hospitality, and Dr. A. Kirillov for 
valuable comments.  
\section*{Appendix: Variant A vs. variant B}
Until Eq.~(4.6) we parallely dealt with variant A ($\epsilon=0$) and 
B ($\epsilon=1$), but beginning from Eq.~(4.7) we simplified by restricting 
to variant A. Now we want to complete the calculation by showing what changes
using variant B. 

The purpose of this appendix is to show in detailed calculations what has
been verbally mentioned in the literature several times, namely the fact 
that differences in solving the factor-ordering problem do not essentially
change the results. For this consideration we now set $\epsilon=1$ in 
Eq.~(4.6) and use the definitions below that equation:
\begin{equation}
  \label{a00}
  \left(\left[v \frac{\partial}{\partial v} v \frac{\partial}{\partial v} 
+  v  \frac{\partial}{\partial v}  
\right] + \gamma^2  v^2 -\Lambda^2    \right) \hat \chi (v,\phi)
 =0.
\end{equation}
The transformation from Eq.~(4.7) to Eq.~(4.8) now leads to 
\begin{equation}
  \label{a01}
  x^2\,\frac{d^2 y}{d x^2}+2x\,\frac{dy}{dx}+(x^2-\Lambda^2)y=0,
\end{equation}
whose solutions are also Bessel functions, modified by a factor $x^{-1/2}$:
\begin{equation}
  \label{a02}
  y(x)=x^{-1/2}\left[\tilde{C}_1 J_{\sqrt{\Lambda^2+1/4}}(x)+\tilde{C}_2 
J_{-\sqrt{\Lambda^2+1/4}}(x) \right]
\end{equation}
where $\tilde{C}_1$, $\tilde{C}_2$ are constants.

This means that the wavefunction (4.17) using variant A is changed by
multiplying
by a factor $\sqrt{3\hbar e^{-3\alpha}/m\phi}$. Moreover, 
the index $\Lambda$ of the Bessel functions must be replaced by
$\sqrt{\Lambda^2+1/4}$.
\section*{References}
\noindent
[1] V. Lukash, H.-J. Schmidt, Astron. Nachr. {\bf 309},  25 (1988); 
H.-J. Schmidt, J. Math. Phys. {\bf  37}, 1244  (1996); gr-qc/9510062.

\noindent
[2] P. Amsterdamski, Phys. Rev. {\bf  D 31}, 3073  (1985).

\noindent
[3] C. Da Silva and R. Williams, Class. Quant. Grav. {\bf  16}, 2681
 (1999).  

\noindent
[4] A. Pelster, 
Dissertation Univ. Stuttgart, Germany, (1996). 

\noindent 
[5] H. Kleinert, Phys. Lett. B {\bf  460}, 36 (1999).

\noindent 
[6] B. DeWitt, Phys. Rev. {\bf 160}, 1113  (1967).

\noindent
[7] L.~P.~Grishchuk and Yu.~V.~Sidorov, p. 700 
in: {\em Proc. 4th Sem. Quantum Gravity},  
eds.:  M. Markov, V. Berezin, and V. Frolov (World Scientific, Singapore,
1988).

\noindent
[8]  D.~N.~Page, p. 82  in: {\em Proc. 5th Sem. Quantum Gravity}, 
eds.: M. Markov, V. Berezin, and V. Frolov (World Scientific, Singapore,
1991).

\noindent
[9] M. Sch\"on and P. Hajicek, Class. Quant. Grav. {\bf 7}, 861
(1990).

\noindent
[10] P. Hajicek, Class. Quant. Grav. {\bf 7}, 871 (1990).

\noindent
[11] N. Landsman, p. 256  in: {\em Current Topics in Mathematical Cosmology}, 
 eds.:  M. Rainer, H.-J. Schmidt (World Scientific, Singapore 1998).

\noindent
[12] S.~P.~Kim, Preprint  gr-qc/9909002 (1999).

\noindent
[13] S. Capozziello, G. Lambiase, {\it Selection rules in minisuperspace 
quantum cosmology},  Preprint (1999), Gen. Relat. Grav. {\bf 32} (2000) in
press. 

\noindent
[14] F. Schunck, E. Mielke, Phys. Rev. {\bf D 50}, 4794  (1994).

\noindent
[15] H.-J. Schmidt, p. 405 in: Differential Geometry and Applications, eds.:
 J. Janyska, D. Krupka  (World Scientific, Singapore, 1990).

\noindent
 [16] S. Odintsov, I. \v Sev\v cenko, Isv. Vuzov Fizika (Univ.
Tomsk), {\bf 7}, 74 (1991);  D. Giulini, {\it What is the geometry of
superspace ?},  
 Preprint gr-qc/9311017 (1993).

\noindent
[17] T. Horiguchi, K. Maeda and M. Sakmaoto, {\it Analysis of the
Wheeler de Witt equation beyond Planck scale}, Preprint hep-th/9409152 (1994).

\noindent
[18] A. Vilenkin, Phys. Rev. {\bf D 50}, 2581 (1994).

\noindent
[19] S. Kim and D. Page, Phys. Rev. {\bf D 45}, R3296 (1992).

\noindent
[20] S. Kim, Phys. Rev. {\bf D 46}, 3403 (1992).

\noindent
[21] U. Bleyer and V. Ivashchuk, Phys. Lett. {\bf B 332}, 292
(1994).

\noindent
[22] D. Page,  in: {\it Quantum concepts in space and time}, p.274 
eds.: C. Isham, R. Penrose  (Oxford Univ. Press 1985).

\noindent
[23] J. Hartle and S. Hawking, Phys. Rev. {\bf D 28}, 2960 
(1983).

\noindent
[24] C.~Kiefer, Phys. Rev. D {\bf 38}, 1761 (1988); H.~Zeh, 
{\it The physical basis of the direction of time} (Springer-Verlag Berlin
1992). 

\noindent
[25] H. Conradi, Phys. Rev. {\bf D 46}, 612 (1992).

\noindent
[26] L. Grishchuk and Ju. Sidorov, J. eksp. i teor. Fiz. {\bf
94}, 29 (1988).

\noindent
[27] L. Amendola, I. Khalatnikov, M. Litterio and F. Occhionero,
Phys. Rev. {\bf D 49}, 1881 (1994).

\noindent
[28] E. Guendelmann, A. Kaganovich, Int. J. Mod. Phys. {\bf D 2},
221 (1993).

\noindent
[29] D. Page, J. Math. Phys. {\bf 32}, 3427 (1991).

\noindent
[30] J. Halliwell, Phys. Rev. {\bf D 38}, 2468 (1988);  H. Kleinert, {\it
Path
 integrals in Quantum Mechanics, Statistics, and
Polymer Physics}, 2nd ed. (World Scientific, Singapore, 1995);  
 M. Jafariztadeh, F. Darabi, A. Rastegar, {\it  On Duru-Kleinert path integral
 in quantum cosmology}, Preprint gr-qc/9811080 (1998);   B. DeWitt, p. 6 in:
 Proc. 8th M. Grossmann Meeting Jerusalem, ed.: T. Piran 
(World Scientific Singapore 1999).
 
\noindent
[31] G. Gibbons and L. Grishchuk, Nucl. Phys. {\bf B 313}, 736 (1989).

\noindent
[32] S. Hawking and D. Page, Nucl. Phys. {\bf B 264}, 185 (1986).

\noindent
[33] V. Melnikov and G. Pevcov, Izv. Vuzov Fizika (Tomsk) {\bf
4}, 45 (1985).

\noindent
[34] S. Reuter and H.-J. Schmidt, p. 243 in: Proc. 5th Intern Conf.
Differential 
Geometry and Appl., eds.: O. Kowalski,
 D. Krupka  (Silesian University Opava 1993). 

\noindent
[35] H.-J. Schmidt, Phys. Rev. {\bf D49}, 6354 (1994), Erratum
 {\bf D54} (1996) 7906.

\noindent
[36] L. Pimentel, O. Obregon, Class. Quant. Grav. {\bf 11} (1994)
2219; J. Fabris, S. Reuter, Gen. Relat. Grav. {\bf 32} (2000) in press.

\noindent
[37] M.~Rainer, Grav. and Cosmol. {\bf 1}, 121 (1995).

\noindent
[38] T. Mongan, Gen. Relat. Grav. {\bf 31}, 1429 (1999).

\noindent
[39] T. Zhang, Y. Shen, Int. J. Theor. Phys. {\bf 38}, 1969   (1999).

\noindent 
[40] A. Peres, Critique of the Wheeler-DeWitt equation, p. 367 in: 
{\it On Einstein's path, Essays in honour of E. Sch\"ucking}, ed.
A. Harvey  (Springer-Verlag Berlin 1999). 

\noindent
[41]  J. Hwang, H. Noh, Preprint 
astro-ph/9908340 (1999), Phys. Rev. D in print.

\noindent
[42] T.~Futamase and K.~Maeda, Phys. Rev. D {\bf 39}, 399 (1989).

\bigskip

\end{document}